\documentclass[reprint,superscriptaddress,amsmath,amssymb,aps,prb]{revtex4-2}
\usepackage{graphicx}
\usepackage{dcolumn}
\usepackage{bm}
\usepackage[colorlinks,allcolors=blue]{hyperref}
\usepackage{natbib}
\usepackage{txfonts}
\usepackage[final]{changes}
\usepackage{changes}
\usepackage{mathrsfs}
\usepackage{amsmath}

\usepackage{physics}
\usepackage{multirow}

\usepackage{orcidlink}
\usepackage{xcolor}
\usepackage{soul}

\begin{document}
\preprint{}
\title{Robust cross-chain surface interstitial electronic states and doping-enhanced superconductivity in 
monolayer $M_2$N ($M$= Ti, Zr, Hf) electrides}

\date{\today}

\author{Da-Bao Zha \orcidlink{0009-0005-7261-5172}}
\affiliation{Laboratory for Quantum Design of Functional Materials, and School of Physics and Electronic Engineering, Jiangsu Normal
University, Xuzhou 221116, China}

\author{Peng Jiang \orcidlink{0000-0002-4291-608X}}
\email{pjiang@jsnu.edu.cn; pjiang93@mail.ustc.edu.cn}
\affiliation{Laboratory for Quantum Design of Functional Materials, and School of Physics and Electronic Engineering, Jiangsu Normal
University, Xuzhou 221116, China}

\author{Yan-Ling Li \orcidlink{0000-0002-0144-5664}}
\email{ylli@jsnu.edu.cn}
\affiliation{Laboratory for Quantum Design of Functional Materials, and School of Physics and Electronic Engineering, Jiangsu Normal
University, Xuzhou 221116, China}

\author{Hai-Qing Lin}
\email{hqlin@zju.edu.cn}
\affiliation{Institute for Advanced Study in Physics and School of Physics, Zhejiang University, Hangzhou 310058, China}

\begin{abstract}

The exploration of electrides holds great promise for advancing both fundamental physics and chemistry, 
owing to their 
unique characteristics 
arising from loosely bound interstitial anionic electrons.
Here we report a class of cross-chain electrides, distinguished by two distinct anionic electron subchannels forming alternating chains in real space. Through structural symmetry analysis and first-principles calculations, we identify two-dimensional $M_2$N ($M$ = Ti, Zr, Hf) materials as prototypical systems exhibiting these unique features. The anionic electron channels on the upper and lower surfaces of these materials display a vertically alternating pattern, with their projected bands revealing momentum-dependent splitting behavior in the reciprocal space, protected by a crystal symmetry operation $\mathcal{O}$. 
Notably, the cross-chain electride characteristic in the $M_2$N monoalyers is 
independent of the layer number and remains robust on the upper 
and lower surfaces of layered structures, presenting pronounced and robust surface interstitial electronic states.
Additionally, we have
explored the superconductivity of these systems, and found that both Ti$_2$N and Zr$_2$N are
intrinsic superconductors with superconducting transition temperatures below 1.0 K. 
Further results show that appropriate hole doping can significantly enhance 
their superconducting transition temperatures and can induce the Hf$_2$N monolayer to exhibit superconductivity. 
Our findings provide valuable insights into the design and tuning of novel electrides 
with enhanced superconducting properties, offering another pathway for 
deeply understanding the interplay between electride behavior and superconductivity in novel materials.

\end{abstract}
%IAEs are confined in the channels formed by cations, which are located in the upper and lower layers respectively, showing a vertical alternating state in the space. 
%From symmetry analysis and lattice model construction, we demonstrate that the cross-type 1D electride are characterized by crystal-rotation 
%symmetries connecting the upper and lower anionic sublattices, which is composed of cationic channels and excess electrons.

\maketitle

\section{Introduction}
Anisotropy in condensed matter systems is critical in determining physical 
properties and has attracted significant attention due to its impact 
on magnetism \cite{PhysRev.105.904,RevModPhys.89.025008,PhysRevX.12.031042,PhysRevX.12.040501}, 
electronic transport \cite{qiao2014high,science.aba5511}, and 
optical behavior \cite{doi:10.1021/acs.nanolett.4c00039,doi:10.1021/acsnano.6b05002}. 
It arises from the breaking of certain symmetries, including time-reversal and spatial 
inversions, and translation operations, which are often associated with various types of ordering, 
such as the atom or lattice arrangement \cite{qiao2014high,zhao2021recent}, 
spin alignment \cite{lado2017origin,wang2019spin,PhysRevB.111.205407}, 
charge distribution \cite{RevModPhys.60.1129,PhysRevLett.123.216403,dye2003electrons}, etc.
Moreover, the interactions between these ordering parameters in the system 
can lead to the emergence of distinct singularities, 
facilitating the discovery of novel and complex physical properties.
A prominent example is altermagnetism or cross-chain antiferromagnet~\cite{PhysRevX.12.031042,PhysRevX.12.040501,ZHANG2025100068}, a novel magnetic phase where momentum-dependent 
spin splitting without spin-orbital coupling emerges as a direct consequence of the 
intricate coupling between spin order and lattice order.
Such an example underscores the importance of investigating the coupling between these ordering parameters, as it not only enhances our understanding of fundamental physical phenomena but also provides critical insights for the discovery of novel material properties. 

\begin{figure}[thp]
	\includegraphics[width=0.9 \linewidth]{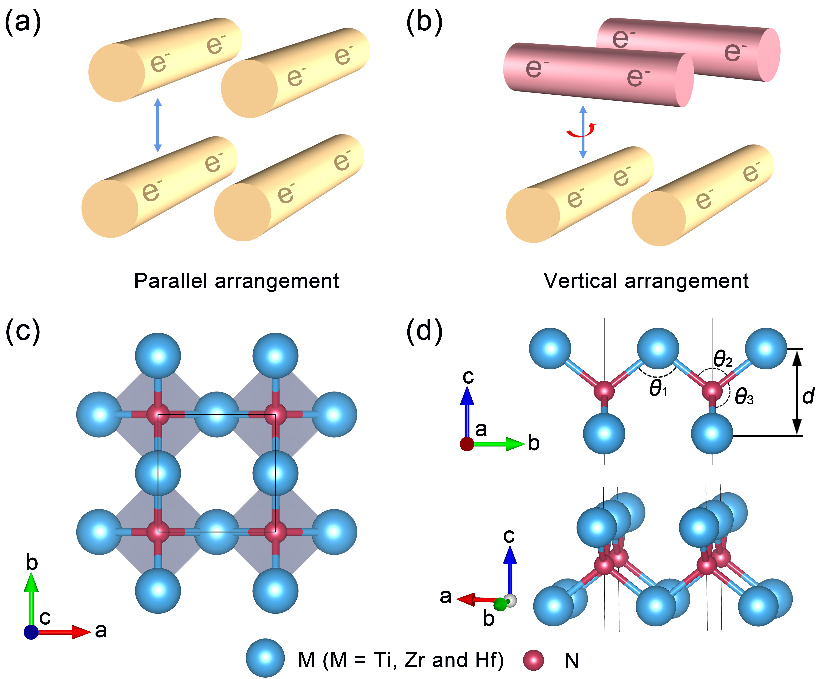}
	\caption{(a) Conventional 1D electride with the parallel-aligned 
	interstitial anionic electron subchannels. (b) Novel cross-chain electrides characterized by 
	alternating interstitial anionic electron subchannels that are aligned parallel 
	within the same plane but intersect in adjacent planes.
	(c) Top and (d) side views of the crystal structures of the $M_2$N monolayers. 
	}\label{fig1}
\end{figure}

Beyond these magnetic states, electrides 
\cite{dye2003electrons,PhysRevB.103.125103,PhysRevLett.127.157002,PhysRevLett.122.097002,PhysRevLett.129.246403}, 
where excess valence electrons are confined to
interstitial spaces in a positively charged crystalline framework and behave as anions, 
constitute an unconventional class of compounds.
The presence of highly mobile non-nucleus-bound electrons in electrides endows them with immense potential for applications in optoelectronics and catalysis \cite{PhysRevLett.123.206402,doi:10.1021/jacs.7b08252}. Based on the connectivity of crystal cavities and channels, as well as the dimensionality of interstitial electron distribution and its coupling with the lattice arrangement, 
electrides can be categorized as zero-, one-, and two-dimensional (0D, 1D, and 2D) phases
\cite{lee2013dicalcium,PhysRevX.4.031023,li2003inorganic,ming2016first}. 
In particular, when the interstitial electron distribution behaves as 
a 1D electron gas, lying between 0D and 2D configurations,
it often gives rise to rich physical phenomena, 
such as superconductivity~\cite{zhang2017electride,doi:10.1021/acs.jpcc.0c00921},  
and nontrivial topological states \cite{PhysRevLett.123.206402,PhysRevLett.120.026401},
due to the restriction of electron motion in a specific direction.
Furthermore, the quantum confinement effect enhances the density of electronic states near the Fermi level, which can induce Stoner-type instability and promote the formation of a magnetic state \cite{sui2019prediction,PhysRevB.110.024413}.
In this case, the 1D electrides can serve as an idea platform to explore the 
connection between electronic anisotropy and lattice structure and also uncovers 
the fundamental mechanisms governing physical properties.
It should be emphasized that, in a conventional 1D electride shown in Fig.~\ref{fig1}(a), 
the interstitial anionic electron (IAE) channels are periodically aligned 
parallel to the nearest neighbors, 
dictated by the translational symmetry of the system.
A natural question is raised: when the IAE channel is regarded as a 
degree of freedom (analogous to a pseudospin), could electrides exhibit 
a cross-chain or $X$-type antiferromagnetic-like state, 
characterized by an arrangement where two distinct IAE subchannels 
are parallel within the same plane but alternate in a crossing pattern between 
adjacent planes [see Fig.~\ref{fig1}(b)]?
Due to the crossing of these IAE channels in electrides with such a stacking order, 
we refer to them as cross-chain electrides.
If such a state were realized, these IAE states might not only exhibit anisotropy in real space but also display distinct signatures in momentum space.
Additionally, numerous studies have shown that electrides provide a great platform for studying superconductivity, yet the role of these anisotropic anionic electron states in superconductivity has not been fully revealed.

In this work, we propose a class of thermodynamically and dynamically stable cross-chain electrides, exemplified by the 2D $M_2$N ($M$ = Ti, Zr and Hf) systems. 
Electronic structure analysis reveals that 1D IAE subchannels 
are anisotropically distributed across the upper and lower surfaces, 
forming a vertically alternating pattern in the real space.
This unique characteristic leads to distinguishable momentum-dependent 
bands for the two anionic electron subchannels, 
highlighting their anisotropic distribution and the significant influence of lattice symmetry.
In addition, the Ti$_2$N and Zr$_2$N monolayers are shown to be
intrinsic superconductors with $T_c$ values of 0.8~K and 0.6~K, respectively. 
Remarkably, the $T_c$ of the Ti$_2$N monolayer can be enhanced to 3.2~K under a doping concentration of 0.6~hole/f.u., and the Hf$_2$N monolayer exhibits a superconducting state under an appropriate 
hole doping, 
providing insights into the suppressive effect of anionic electrons in electrides on superconductivity.

\section{Computational Methods and Details}
The calculations for structural and electronic properties 
of the $M_2$N ($M$ = Ti, Zr and Hf) monolayers are carried out 
using the Vienna Ab-initio Simulation Package (VASP) \cite{kresse1996efficient,kresse1999ultrasoft}, 
which is based on density functional theory (DFT) within projector augmented wave (PAW) method \cite{blochl1994projector}.
The generalized gradient approximation with the Perdew-Burke-Ernzerhof (PBE) functional is used to describe
the exchange-correlation term \cite{PhysRevLett.77.3865}. The cutoff energy of the plane wave 
is set to 500 eV. All crystal structures are relaxed until the forces per 
atom are less than 1 meV/{\AA} and total energies are converged to $10^{-6}$ eV. 
The $k$-mesh grids of 14$\times$14$\times$1 and 14$\times$14$\times$10 are 
sampled in the Brillouin zone (BZ)
by using the Monkhorst-Pack scheme in the monolayer and bulk $M_2$N systems, respectively.
For the multilayer systems, the interlayer van der Waals (vdW) interaction is
described by the DFT-D3 method \cite{10.1063/1.3382344}.
The electron localization function (ELF) is used to analyze 
the inter-atomic bonding \cite{10.1063/1.458517}.
The phonon spectra and electron-phonon coupling (EPC) calculations are 
performed using the density functional perturbation theory (DFPT), 
as implemented in the QUANTUM ESPRESSO (QE) package \cite{Giannozzi_2009},
in which the $k$ mesh and $q$ mesh are both set to 8$\times$8$\times$1.
The electronic states are described using optimized norm-conserving 
Vanderbilt (ONCV) pseudopotentials \cite{PhysRevB.88.085117}, with cutoff energies 
of 90~Ry for the wave function and 450~Ry for the charge density.
In addition, we have calculated the phonon spectra using the frozen-phonon method 
implemented in VASP, employing a 5$\times$5$\times$1 supercell, in order to cross-validate 
the dynamical stability results obtained from both DFPT and frozen-phonon approaches.
The superconducting $T_c$ is calculated by using the 
McMillan-Allen-Dynes formula \cite{PhysRevB.12.905}, 
namely, $T_\mathrm{c}=\frac{\omega_\mathrm{log}}{1.2} \exp\biggl[-\frac{1.04(1+\lambda)}{\lambda-\mu^{*}(1+0.62\lambda)}\biggr]$,
where a typical value of 0.11 for the effective Coulomb repulsion parameter $\mu^*$ is adopted \cite{PhysRevB.99.220503}.

\section{Results and Discussion}
\subsection{Structural Characteristics and Stability}
%%%%%
The optimized structures of the monolayer transition metal nitrides $M_2$N ($M$ = Ti, Zr, and Hf) 
are presented in Figs.~\ref{fig1}(c) and \ref{fig1}(d). 
Similar to 2D SnO$_2$ \cite{PhysRevB.102.195408}, 
the $M_2$N monolayers hold a tetrahedral lattice with a buckled geometrical configuration 
consisting of $M$N$_4$ tetrahedron. As shown in Fig.~\ref{fig1}(d), 
$M$-N-$M$-N-$M$ atoms in the 
$M_2$N monolayers form an open dome-shaped channel along the $x$ and $y$ directions, respectively. 
These two open channels can be connected by $S_{4z}$ 
operation since the $M_2$N monolayers 
belong to $P\overline{4}m2$ (no.~115) space group with $D_{2d}$ point symmetry.
Note that all $M_2$N systems share similar crystal structures and electronic 
band characteristics (as discussed below), we thus primarily focus on Ti$_2$N as 
a representative example in the following analysis. 
We perform density functional theory calculations and 
find that the in-plane lattice constants
are $a$ = $b$ = 2.99~{\AA}. 
Other structural parameters are summarized 
in Table~\ref{table1}.

\begin{table}[htbp]
    \centering
    \caption{Calculated structural parameters of the $M_2$N ($M$ = Ti, Zr and Hf) monolayers,
    in which $a$, $d$ and $d_\mathrm{M-N}$ denote the lattice constant, the buckled height, 
    and the $M$-N bond length, respectively. 
    $\theta_1$, $\theta_2$, and $\theta_3$ are the bond angles.}\label{table1}
    \setlength{\tabcolsep}{1.3mm}{
    \renewcommand{\arraystretch}{1.5}{
    \begin{tabular}{*{7}{c}} %
    \hline
    \hline
    \multirow{2}{*}{$M_2$N} & \multirow{2}{*}{$a = b$ (\AA)} & \multirow{2}{*}{$d$ (\AA)} & \multirow{2}{*}{$d_\mathrm{M-N}$ (\AA)} & \multicolumn{3}{c}{bond angles ($^\circ$)} \\
    & & & & $\theta_1$ & $\theta_2$ & $\theta_3$ \\
    \hline
    Ti$_2$N & 2.998 & 2.396 & 1.920 & 102.70 & 102.70 & 112.90 \\
    Zr$_2$N & 3.248 & 2.561 & 2.068 & 103.50 & 103.50 & 112.54 \\
    Hf$_2$N & 3.196 & 2.574 & 2.052 & 102.31 & 102.31 & 113.17 \\
    \hline
    \hline
    \end{tabular}}}
\end{table}

 \begin{figure*}[tbhp]
         \centering
        \includegraphics[width=0.8 \linewidth]{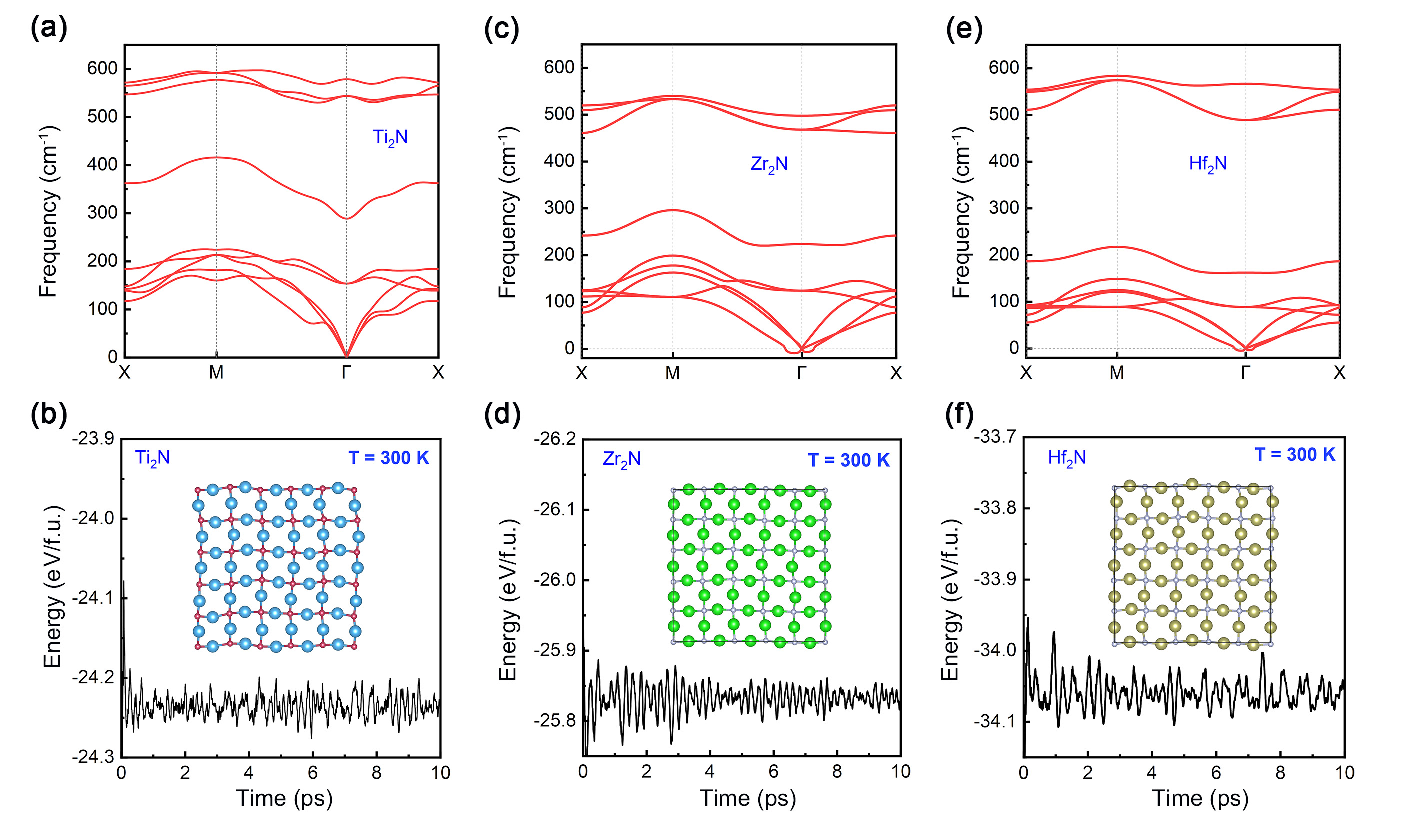}
	 \caption{Phonon spectra and AIMD simulation results at 300~K of (a), (b) Ti$_2$N, 
	 (c), (d) Zr$_2$N, and (e), (f) Hf$_2$N monolayers, 
         respectively. The insets show the structural snapshots of such $M_2$N monolayers at 10~ps.
		}\label{fig2}
\end{figure*}

To evaluate the energetic stability of the Ti$_2$N monolayer, we calculate its formation 
energy defined as $E_{for}=(E_\mathrm{Ti_2N}-2E_\mathrm{Ti}-E_\mathrm{N})$/3, where 
$E_\mathrm{Ti_2N}$ are the total energy of the Ti$_2$N monolayer; 
$E_\mathrm{Ti}$ and $E_\mathrm{N}$ represent the atomic chemical potentials 
derived from their respective stable phases under ambient conditions. 
The calculated $E_{for}$ is about -0.38 eV/atom, indicating that the monolayer is energetically favorable for experimental synthesis. 
    As shown in Fig.~S1 of the Supplemental Material (SM) \cite{supp}, we calculated a cleavage energy of 
4.28~J/m$^2$, exceeding that of typical vdW materials like graphite (~0.37~J/m$^2$ \cite{wang2015measurement}). 
This higher value arises from strong interlayer coupling via delocalized IAEs (discussed later), 
enhancing cohesion beyond conventional vdW interactions. Despite the increased exfoliation energy, 
the value remains feasible for experimental monolayer realization, as supported by recent 
mechanical exfoliation of non-vdW materials \cite{jiang2023mechanical}.
The phonon dispersion obtained by QE package is 
then calculated and the absence of imaginary frequencies 
confirms the dynamical stability of the Ti$_2$N monolayer [see Fig.~\ref{fig2}(a)]. 
Specifically, we performed frozen-phonon calculations using the finite displacement method
and the resulting phonon spectra (see Fig.~S2 in the SM \cite{supp}) shows no significant 
imaginary frequencies across the Brillouin zone, except for negligible 
imaginary modes around the $\Gamma$ point, a common feature in 2D systems, 
 thus verifying the dynamical stability of the Ti$_2$N monolayer.
Moreover, the thermal stability is assessed via {\it ab-initio} molecular 
dynamics simulations [see Fig.~\ref{fig2}(b)]. 
The results indicate that the Ti$_2$N monolayer exhibits 
robust thermal stability, with stable energy fluctuations 
and no significant structural reconstruction.

 \begin{figure}[thp]
         \centering
        \includegraphics[width=0.96 \linewidth]{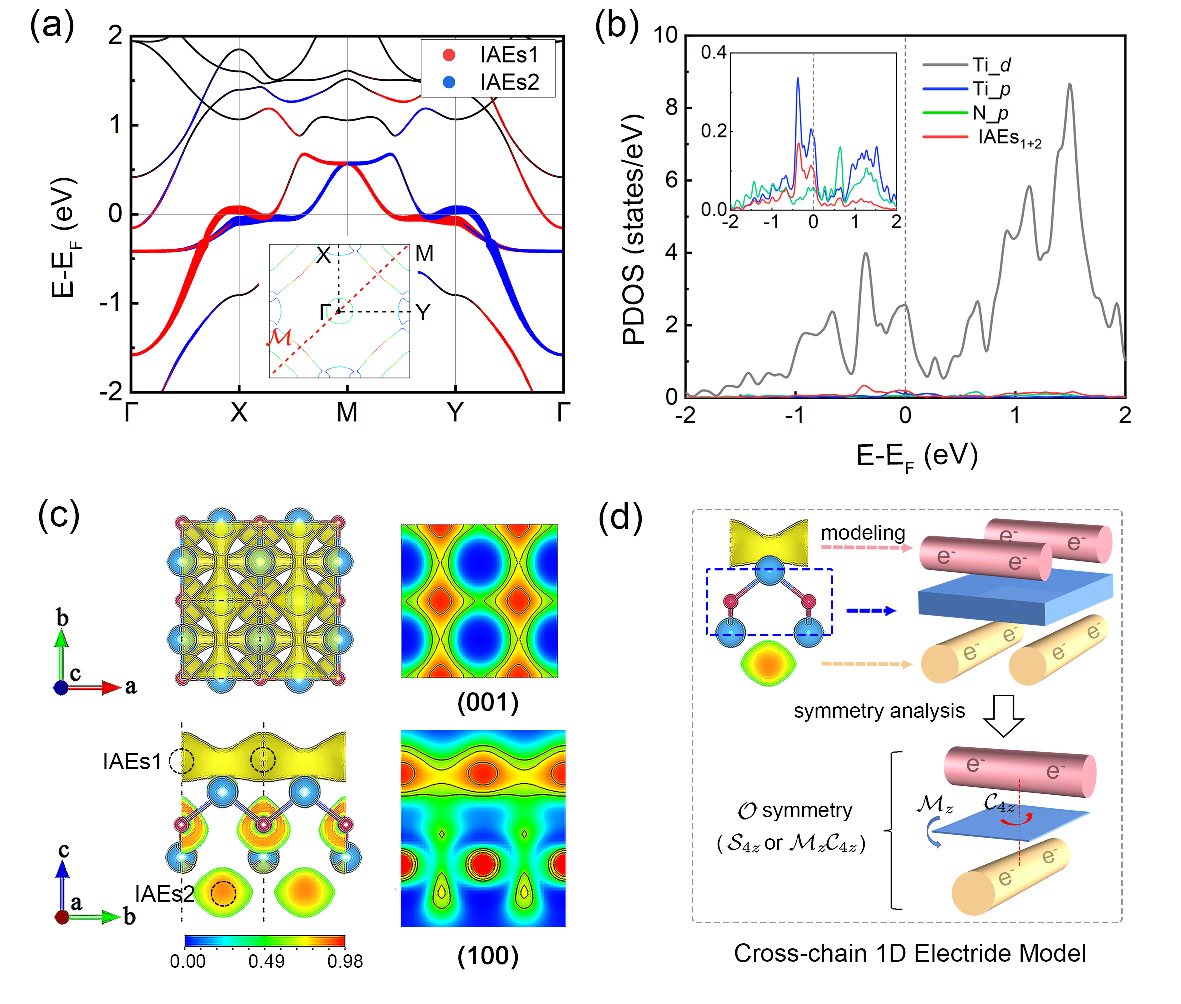}
	\caption{(a) IAE-projected band structure of the Ti$_2$N monolayer, and the
     inset shows the Fermi surface in the first BZ.
     (b) PDOS of the Ti$_2$N monolayer, the inset displays the enlarged view of 
     the contribution of interstitial electrons around the Fermi level. 
     (c) Different views of the ELF maps with the isosurface values of 0.55.
     The black dashed circles denote the position of pseudoatoms. 
     %(d) ELF of 1-, 2-, 3-, 4-layer and bulk Ti$_2$N structures.
         (d) The symmetry analysis of the cross-chain 1D electride model, 
     where the cylinder and cuboid represents IAEs and 2D atomic layer, respectively.
	}\label{fig3}
\end{figure}

\subsection{Momentum-dependent interstitial anionic electron distribution}
%\textbf{Momentum-dependent interstitial anionic electron distribution}.

To further reveal the electronic properties of the Ti$_2$N monolayer, 
we analyze its electronic band structure and the corresponding projected density of states (PDOS).
As shown in Fig.~\ref{fig3}, there are two 
hole-like bands and one electron-like band crossing the Fermi
level, with no intersections between them [see the
inset in Fig.~\ref{fig3}(a)], indicating the semimetallic characteristic of the Ti$_2$N monolayer.
The PDOS shown in Fig.~\ref{fig3}(b) demonstrates that these bands are predominantly derived from 
Ti-$d$ orbitals, while contributions from Ti-$p$ and N-$p$ orbitals are minimal.
Given the usual valence states of
Ti (Ti$^{2+}$, Ti$^{3+}$, or Ti$^{4+}$) in the nitrides or oxides
and N (N$^{3-}$) in the metal nitrides, it is naturally expected that
the system may possess excess electrons, behaving as IAEs.
This electronic behavior can be confirmed by the ELF that is usually 
used to characterize the bonding feature \cite{PhysRevB.104.035430,PhysRevB.111.205426}
and the corresponding results are displayed 
in Fig.~\ref{fig3}(c). 
We can see that two distinct excess electronic states are loosely confined to the 
dome-shaped interstitial spaces above and below the atomic layers, 
with the upper and lower IAEs labeled as IAEs1 and IAEs2, 
respectively.
Notably, both IAEs1 and IAEs2 form continuous 1D electronic states 
with spatially orthogonal and separated distributions, 
leading to an alternating electronic configuration in real space. 
Furthermore, to validate the reliability of our results 
and reduce potential self-interaction errors, we additionally performed hybrid functional (HSE06) calculations for the Ti$_2$N monolayer \cite{10.1063/1.2404663}. The key electronic 
distributions and features associated with the 
IAE states remain qualitatively unchanged compared to the PBE results, 
confirming the robustness of our conclusions with respect to the choice of exchange–correlation functional.

In order to estimate the number of interstitial electrons in the Ti$_2$N
monolayer, we have performed a Bader charge analysis, which reveals that 
the Bader charges of Ti, N, and IAEs are 1.77~$\lvert$e$\rvert$, 
-1.53~$\lvert$e$\rvert$, and -2.00~$\lvert$e$\rvert$, respectively. 
These indicate that interstitial charges function 
as anionic electrons and are primarily derived from Ti atoms. 
Combined with the above results, the chemical formula of 
the Ti$_2$N electride can be expressed as (Ti$_2$N)$^{2+}$ $\cdot$2$e^{-}$. 
Additionally, we employ a pseudoatom method to calculate the IAE-projected band structure and PDOS, 
as shown in Figs.~\ref{fig3}(a) and~\ref{fig3}(b).
It is emphasized that this method is commonly and broadly adopted in electrides 
to extract IAE-related properties without 
altering the underlying electronic structure \cite{PhysRevB.106.L060506,li2021electron,doi:10.1021/jacs.5b00242,PhysRevMaterials.7.114805}.
It is shown that IAEs1 and IAEs2 exhibit a distinct energy-splitting feature, 
and are related by a diagonal mirror symmetry $\mathcal{M}$ in the 
momentum ($\boldsymbol k$) space, 
similar to previously reported spin-split band structures for altermagnetic and cross-chain antiferromagnetic phases \cite{PhysRevX.12.031042,PhysRevX.12.040501,ZHANG2025100068,PhysRevB.110.174429}. 

Next, we analyze symmetry conditions for the distinguishable behavior of
$\boldsymbol k$-dependent IAE distribution.
Without loss of generality, we consider the Ti$_2$N system with 
two sublattices: the top sublattice with $z >$0 and the bottom one with $z <$ 0, in which 
$z = 0$ corresponds to the N atomic layer [see Fig.~\ref{fig3}(d)]. 
By definition, the electron charge of 
the top/bottom sublattice can be expressed simply as
$\rho_{1/2}$= $e \sum_{n} \int_{\Omega} \int_{z>0/z<0} f_{n \boldsymbol k} \braket{\psi_{n \boldsymbol k}}{\psi_{n \boldsymbol k}}$ $d\boldsymbol r d\boldsymbol k$
= $\int_{\Omega} \rho_{1/2}(\boldsymbol k) d \boldsymbol k$,
in which the numbers 1 and 2 denote the layer index for
convenience, $e$ is the electron charge,
$f_{n \boldsymbol k}$ is the Fermi-Dirac distribution function,
$n$ is the band index for the occupied states,
and {\it \textbf{k}} = ($k_x$,$k_y$) indicate the momentum in the 2D BZ.
Under the $\mathcal{O}$ symmetry operation, {\it \textbf{k}} should 
be changed to $\boldsymbol k^{\prime}$ = ($k_y$,$k_x$), namely, 
$\mathcal{O}\boldsymbol k$ = $\boldsymbol k^{\prime}$.
It is clear that $\rho_{1}$ and $\rho_{2}$ can be connected by 
$\mathcal{O}$ operation ($\mathcal{S}_{4z}$ or
$\mathcal{C}_{4z}\mathcal{M}_{z}$) from the structural symmetry (see Fig.~\ref{fig3}(d)), that is, 
$\mathcal{O}^\dagger \rho_{1} \mathcal{O}$=$\rho_{2}$. 
Thus, one can get 
\begin{equation}
    \begin{split}
    \mathcal{O}^\dagger \varepsilon_{1n}(\boldsymbol k) \mathcal{O}
     = \mathcal{O}^\dagger \int_{z>0} \bra{\psi_{n \boldsymbol k}(\boldsymbol r)} \hat{H_1}
    \ket{\psi_{n \boldsymbol k}(\boldsymbol r)} d{\boldsymbol k}~
    \mathcal{O} \\
        =\int_{z<0} \bra{\psi_{n \mathcal{O} \boldsymbol k}(\boldsymbol r)} \hat{H_2}
        \ket{\psi_{n \mathcal{O} \boldsymbol k}(\boldsymbol r)} d{\boldsymbol r}
        =\varepsilon_{2n}({\boldsymbol k}^{\prime}).
    \end{split}
\end{equation}
Meanwhile, the broken $\mathcal{M}_{z}$ symmetry 
in the $M_2$N structures can guarantee that 
$\varepsilon_{1n}$($\boldsymbol k$) $\ne$ $\varepsilon_{2n}$($\boldsymbol k$) is
occurred, meaning that the energy distributions of IAEs1 and IAEs2 are 
always de-degenerate at the same $\boldsymbol k$ path 
but are symmetric through the diagonal mirror symmetry of 2D BZ. 

 \begin{figure}[thp]
         \centering
        \includegraphics[width=0.96 \linewidth]{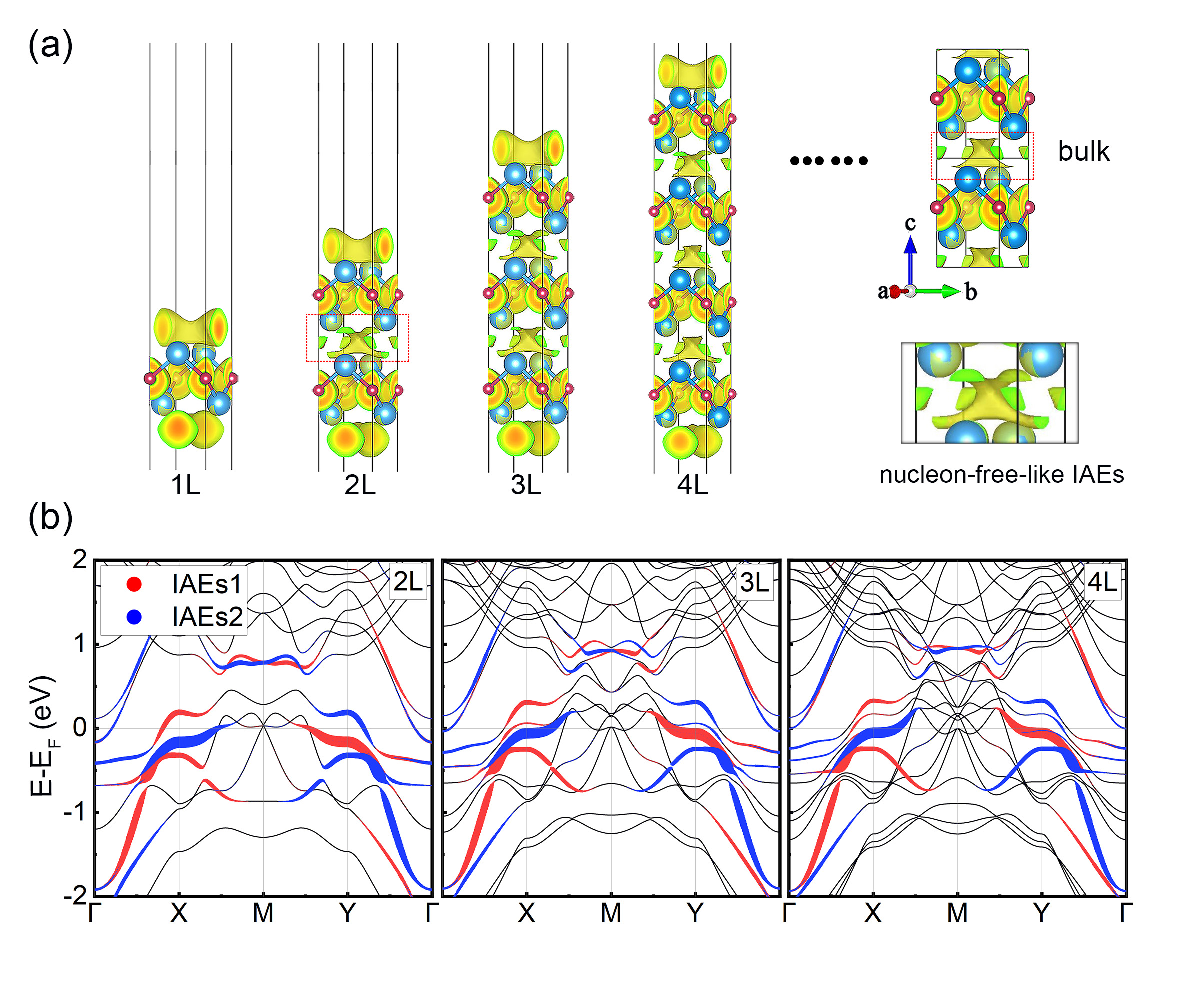}
	\caption{ Layer-dependent (a) ELF maps with the isosurface values of 0.55
	and (b) band structures of the layered Ti$_2$N structures.
	}\label{fig4}
\end{figure}

To examine the layer-dependent distribution of IAEs, 
ELF calculations for one-, two-, three-layer, four-layer, and bulk Ti$_2$N are conducted [see Fig.~\ref{fig4}(a)]. 
It is indicated that the alternating IAE configuration persists in multi-layered systems 
and remains spatially and perpendicularly separated on the upper and lower 
surfaces of 2D layered materials. 
Moreover, the momentum-dependent IAE characteristics persist in the band structures of multilayer 
systems [see Fig.~\ref{fig4}(b)], manifesting a stable surface distribution of IAEs
 at the top and bottom surfaces.
This surface-confinement behavior 
here originates from quantum confinement and the symmetry-protected localization of IAEs.
In addition, the multilayer Ti$_2$N exhibits a nucleon-free-like IAE behavior between its layers, akin to the bulk Ti$_2$N electride.

\begin{figure}[thp]
	\includegraphics[width=0.95 \linewidth]{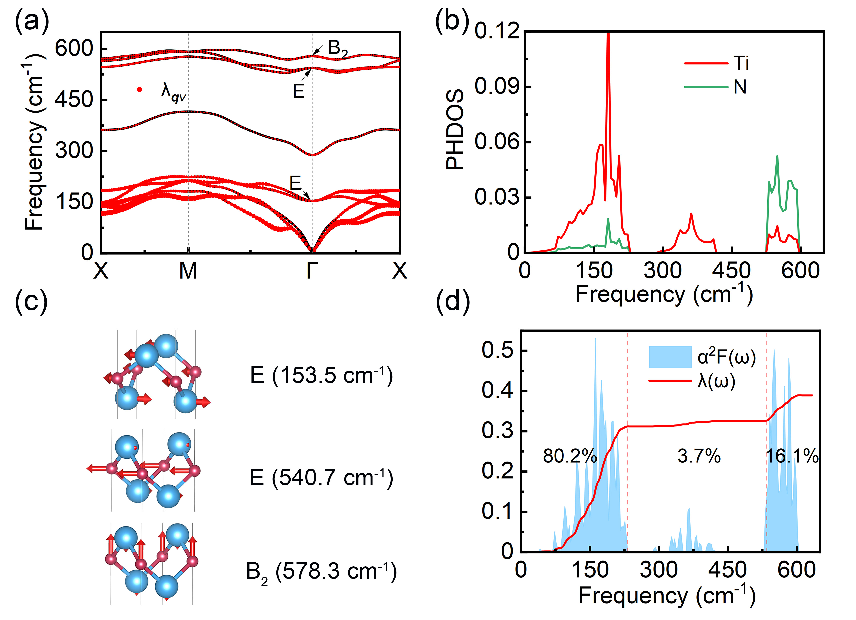}
	\caption{(a) Phonon spectrum with the EPC strength $\lambda_{\boldsymbol{q}\nu}$ for each 
	phonon mode and (b) atomic projected PHDOS of the Ti$_2$N monolayer. 
	The size of dots is proportional to the
    $\lambda_{\boldsymbol{q}\nu}$ strength.
    (c) Eigenvectors of the infrared and Raman (I+R) active modes at the $\Gamma$ point.
    (d) Eliashberg spectral function $\alpha^2F(\omega)$, represented by the shaded area,
    and the cumulative frequency-dependent EPC constant $\lambda(\omega)$ represented
    by the red line of the Ti$_2$N monolayer. 
	}\label{fig5}
\end{figure}

\subsection{Enhanced superconducting properties by the hole-doping}
%\textbf{Enhanced superconducting properties by the hole-doping}.

To gain a deeper understanding into EPC properties of the Ti$_2$N monolayer, 
the phonon spectrum with the EPC strength for each
phonon mode $\lambda_{\boldsymbol{q}\nu}$, atomic projected phonon DOS (PHDOS),
and Eliashberg spectral function $\alpha^2F(\omega)$ are calculated and the
results are shown in Fig.~\ref{fig5}. 
Before discussing the EPC properties, it should be emphasized that the electronic band structures 
obtained from QE are in excellent agreement with those from VASP (see Fig.~S3 in the SM \cite{supp}). 
This consistency provides a solid methodological basis for subsequently 
using QE to calculate EPC and superconducting properties.
From the phonon spectrum and PHDOS present in
Figs.~\ref{fig5}(a) and \ref{fig5}(b), 
we can find that the main contribution to the EPC strength
attributes to the low-frequency phonon modes below 250~cm$^{-1}$,
which are primarily associated with the stretching vibrations between Ti and N 
atoms in the infrared and Raman $E$ modes [see the vibration 
pattern at a frequency of 153.5 cm$^{-1}$ in Fig.~\ref{fig5}(c)].
These low-frequency vibrations contribute 80.2\% of the total
EPC strength, as shown in Fig.~\ref{fig5}(d).
Furthermore, the high-frequency phonon modes above 500~cm$^{-1}$ mainly
originate from the in-plane and out-of-plane vibrations of N atoms, which
contribute 16.1\% of the total EPC strength.
In contrast, the intermediate-frequency phonon modes between 250 and 500 cm$^{-1}$ 
make a negligible contribution.
Therefore, the main contribution to the EPC strength is from Ti vibrations.
Using the McMillan-Allen-Dynes formula \cite{PhysRevB.12.905}, 
the superconducting $T_c$ of the Ti$_2$N monolayer is estimated to be around 0.8~K.
We have also analyzed the superconductivity of the bulk Ti$_2$N and  found its $T_c$ 
to be 5.5~K, significantly higher than that of the Ti$_2$N monolayer. 
The increase of $T_c$ is close to the delocalization of IAEs with increasing
layer number.

\begin{figure}[thp]
	\includegraphics[width=0.95 \linewidth]{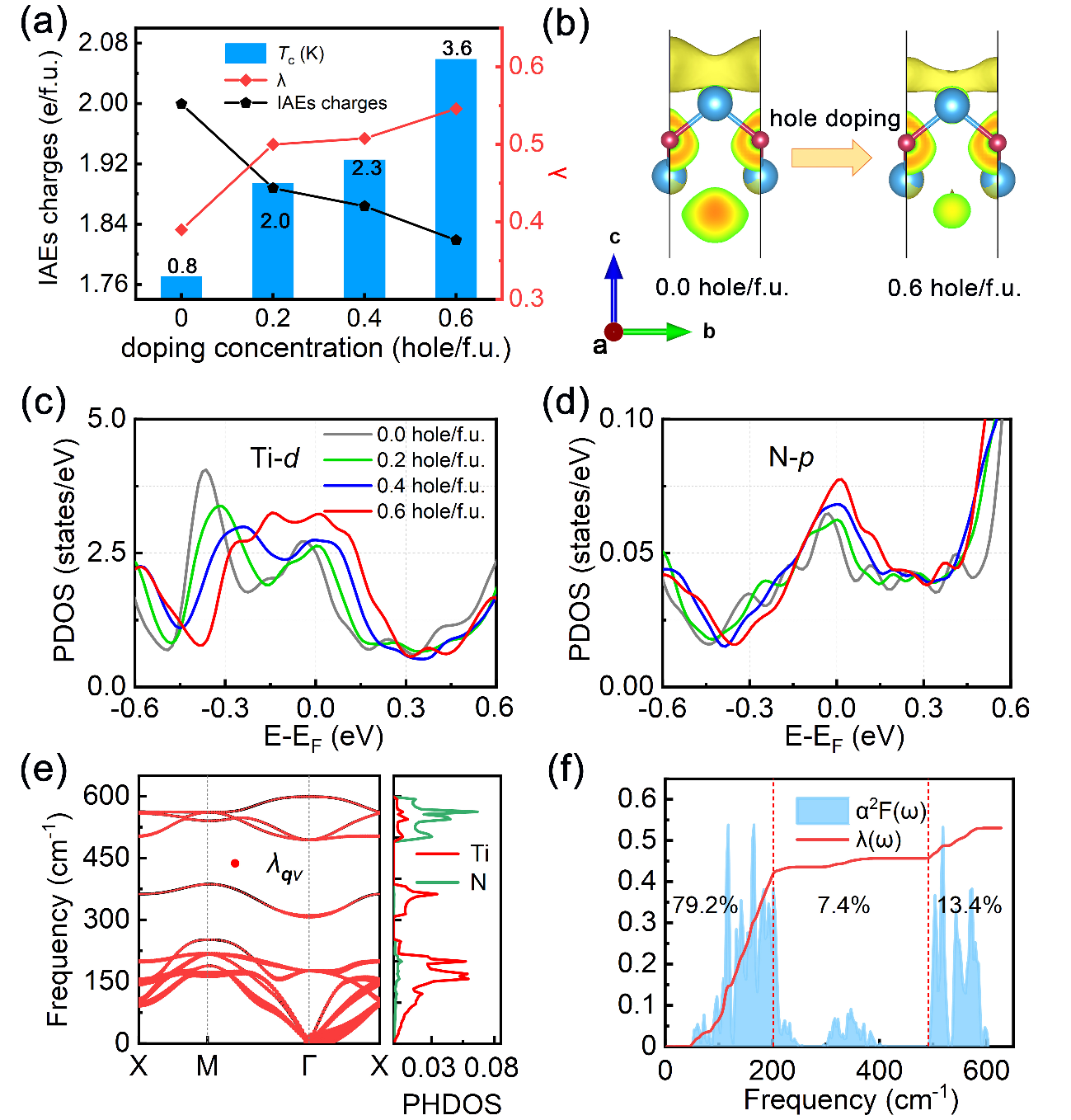}
    \caption{(a) Hole doping concentration ($n_h$) dependent $T_c$ and $\lambda$ in the Ti$_2$N monolayer. 
    (b) ELF evolution from $n_h$ = 0.0~hole/f.u. to $n_h$ = 0.6~hole/f.u., and 
    the isosurface values are set to 0.55.
    (c) Ti-$d$ and (d) N-$p$ orbitals PDOS at different $n_h$. 
    (e) Phonon spectrum with the $\gamma_{\boldsymbol{q}\nu}$ attached, atom projected PHDOS, 
    and (f) Eliashberg spectral function
    $\alpha^2F(\omega)$ and cumulative frequency-dependent EPC constant of the Ti$_2$N monolayer at $n_h$ = 0.6~hole/f.u.. 
	}\label{fig6}
\end{figure}

To investigate the influence of changes in IAEs on the electronic structure of
 (Ti$_2$N)$^{2+}$ $\cdot$2$e^{-}$, 
we introduce hole carriers to neutralize anionic electrons,
thereby modulating the electronic structure.
First, the $T_c$ at different doping concentrations
of hole carriers ($n_h$) is examined, as depicted in Fig.~\ref{fig6}(a). 
With increasing $n_h$, a significant rise in the
$T_\mathrm{c}$ has observed. More specifically, hole doping increases 
the EPC strength from $\lambda$ = 0.39 in the undoped case 
to $\lambda$ = 0.55 at $n_h$ = 0.6 hole/f.u.
It is well known that the $T_\mathrm{c}$ is closely 
related to $\lambda$ 
and the logarithmical averaged frequency $\omega_\mathrm{log}$. 
Note that, although $\omega_\mathrm{log}$ is decreased with
increasing $n_h$, the enhancement of $\lambda$ dominates an increase in the $T_\mathrm{c}$. 
When $n_h$ reaches 0.6~hole/f.u., the
$T_\mathrm{c}$ rises from 0.8~K in the pristine case to 3.2 K. 
Although the $T_\mathrm{c}$ remains relatively low at moderate hole doping concentration, 
it increases fourfold compared to the undoped case, which can be attributed 
to the neutralization of anionic electrons.
This trend is clearly confirmed by the ELF shown in Fig.~\ref{fig6}(b). 
It is found that the spatial distribution of IAEs becomes more confined, 
indicating that the number of IAEs decreases with increasing $n_h$. 
Consequently, the Bader charge of IAEs decreases from -2.00~$\lvert$e$\rvert$ 
to -1.80~$\lvert$e$\rvert$ when $n_h$ increases from 0.0~hole/f.u. to 0.6~hole/f.u.

As illustrated in Figs.~\ref{fig6}(c) and ~\ref{fig6}(d), 
a gradual reduction in the number of IAEs on the surface leads 
to an increase in delocalized electrons near the Fermi surface, 
predominantly originating from Ti-$d$ and N-$p$ orbitals. 
From the electronic perspective, the increase of the doping concentration 
reduces the Coulomb attraction between IAEs and the host cationic lattice, 
facilitating Cooper pair formation and enhancing superconducting properties \cite{ChinPhysLett.40.107401}.
It should be emphasized that hole doping can be effectively achieved by tuning the type of charge carriers, as experimentally validated via electrostatic gate modulation techniques~\cite{jiang2018controlling,wu2023electrostatic}.
This approach enables carrier accumulation or depletion at the surface or interface without 
introducing significant structural distortion to the monolayer system.
In this context, we investigate the effect of hole doping on phononic properties, 
using the representative case of 0.6~hole/f.u.~as an example.
The phonon dispersion with the attached $\lambda_{\boldsymbol{q}\nu}$ and 
atomic-projected PHDOS for the Ti$_2$N monolayer at 0.6~hole/f.u.~are shown in Fig.~\ref{fig6}(e).
Similar to the undoped case, the high-frequency region exhibits significant phonon linewidth contributions, primarily associated with the vibrations of N atoms. 
Meanwhile, Figure~\ref{fig6}(f) reveals an increase in the proportion of 
$\lambda$ in the low- and intermediate-frequency regions.
From the above results, it can be concluded that hole doping induces two key effects: 
first, the reduction in the number of excess electrons enhances the density of delocalized 
electrons near the Fermi surface; second, the weakening of Coulomb interactions 
between IAEs and host cations facilitates favorable vibrations of Ti atoms 
in the low- and intermediate-frequency regions of phonons, 
thereby increasing their contribution to superconductivity.

\begin{figure}[thp]
	\includegraphics[width=0.85 \linewidth]{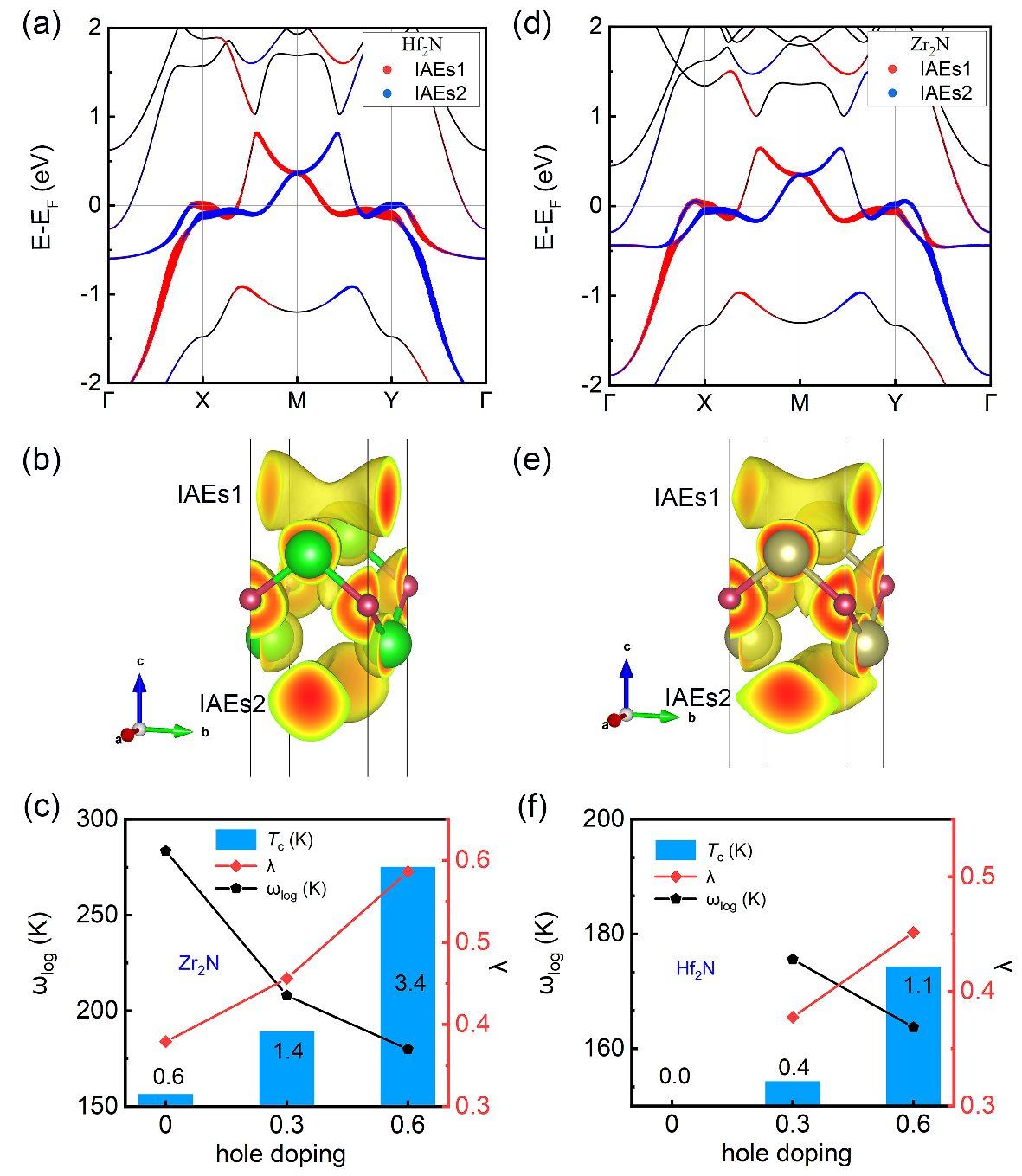}
    \caption{IAE-projected band structure and ELF maps of the (a), (b) Zr$_2$N and (d), (e) Hf$_2$N monolayers.
    The ELF isosurface values are set to 0.55.
Hole doping concentration ($n_h$) dependent $T_c$, $\omega_\mathrm{log}$, and $\lambda$ of
 the (c) Zr$_2$N and (f) Hf$_2$N monolayers.    
	}\label{fig7}
\end{figure}

Considering that Zr and Hf belong to the same group as titanium, we substituted Ti 
atoms in Ti$_2$N with Zr and Hf, resulting in Zr$_2$N and Hf$_2$N systems, 
both of which are dynamically stable [see Figs.~\ref{fig2}(c) -\ref{fig2}(f)]. 
Their electronic structures were subsequently analyzed, as shown in 
Figs.~\ref{fig7}(a), \ref{fig7}(b) and ~\ref{fig7}(d), \ref{fig7}(e), 
revealing that Zr$_2$N and Hf$_2$N monolayers exhibit cross-chain 
1D electride features. This behavior can be attributed to their identical 
crystal structure to Ti$_2$N and the shared valence state of Zr and Hf.
Furthermore, the superconducting properties of Zr$_2$N and 
Hf$_2$N monolayers are investigated, with results presented in Figs.~\ref{fig7}(c) and \ref{fig7}(f). 
It is found that the Zr$_2$N monolayer is a intrinsic superconductor with a $T_c$ of 
0.6~K in the undoped state, while the Hf$_2$N monolayer shows no superconductivity. 
Interestingly, with hole doping, both of the two systems demonstrate enhanced 
superconducting behavior. The $T_c$ of the Zr$_2$N monolayer is increased significantly 
with higher doping concentrations. Moreover, the Hf$_2$N monolayer 
exhibits a superconducting state under the hole doping, and its $T_c$ continues to rise 
with increasing doping concentration. These findings support the conclusion 
that the introduction of the holes in the cross-chain electrides $M_2$N system can effectively 
enhance their superconductivity by strengthening the Coulomb interaction 
between delocalized IAEs and host cations.

\section{Conclusion}
In summary, we have proposed a class of cross-chain one-dimensional (1D) electrides, 
exemplified by the theoretically predicted 2D $M_2$N ($M$ = Ti, Zr, and Hf) monolayers.  
These materials exhibit unique anisotropic IAE distributions, 
characterized by vertically alternating 1D chains along surface atomic channels. 
Symmetry analysis reveals that the distinct IAE subchannels on the upper and lower 
surfaces give rise to momentum-dependent subchannel-splitting features in the 
electronic band structure. Remarkably, this cross-chain electride characteristic 
is retained regardless of the number of layers, owing to the imposed $S_{4z}$ symmetry 
of the atomic framework. As a result, these systems host robust cross-chain 
surface interstitial electronic states.
Importantly, Ti$_2$N and Zr$_2$N monolayers are predicted to be intrinsic superconductors, 
with relatively low superconducting $T_c$.
Furthermore, the hole doping is demonstrated to enhance the EPC and $T_c$ in these electrides, 
driven by an increase in electronic states near 
the Fermi level and a strengthened interaction between the 
host cations and delocalized IAEs.
Our findings not only report the cross-chain 1D electrides with superconductivity, but also demonstrate the significant role of the coupling of lattice and charge orderings 
between in shaping the physical properties in electrides, offering a platform to explore the interaction 
between electride state and superconductivity.

\section*{acknowledgments}
This work is funded by the National Natural Science Foundation 
of China (Grants No.~12204202, No.~12074153, and No.~12088101), 
the Basic Research Program of Jiangsu (Grant No.~BK20220679), 
and the Natural Science Fund for Colleges and Universities in Jiangsu 
Province (Grant No.~22KJB140010).

\bibliography{ref}

\end{document}